\documentclass[12pt,preprint2]{emulateapj}
\usepackage{pslatex}
\usepackage[T1]{fontenc}
\usepackage[latin1]{inputenc}
\usepackage{subfigure}
\usepackage{amsmath}
\usepackage{graphicx,graphics}
\usepackage{amssymb}
\usepackage[english]{babel}

\newcommand{\gap}{\;\rlap{\lower 2.5pt \hbox{$\sim$}}\raise 1.5pt\hbox{$>$}\;}
\newcommand{\lap}{\;\rlap{\lower 2.5pt \hbox{$\sim$}}\raise 1.5pt\hbox{$<$}\;}
\newcommand{\beq}{\begin{equation}}
\newcommand{\eeq}{\end{equation}}

\newcommand{\msun}{M_\odot}

\shorttitle{Cusp Generation}
\shortauthors{Merritt \& Szell}
\begin{document}

\title{Dynamical Cusp Regeneration}

\author{David Merritt and Andras Szell}
\affil{Department of Physics, Rochester Institute of Technology,
Rochester, NY 14623}

\begin{abstract}
After being destroyed by a binary supermassive black hole,
a stellar density cusp can regrow at the center of a galaxy
via energy exchange between stars moving in the gravitational 
field of the single, coalesced hole.
We illustrate this process via high-accuracy $N$-body simulations.
Regeneration requires roughly one relaxation time
and the new cusp extends to a distance of roughly one-fifth the
black hole's influence radius,
with density $\rho \sim r^{-7/4}$; the mass in the cusp is
of order 10\% the mass of the black hole.
Growth of the cusp is preceded by a stage in which the
stellar velocity dispersion evolves toward isotropy
and away from the tangentially-anisotropic state 
induced by the binary.
We show that density profiles similar to those observed
at the center of the Milky Way and M32 
can regenerate themselves in several Gyr following infall
of a second black hole; the presence of density cusps at the 
centers of these galaxies can therefore not be used to infer that no 
merger has occurred.
We argue that $\rho\sim r^{-7/4}$ density cusps are ubiquitous
in stellar spheroids fainter than $M_V\approx -18.5$ that contain
supermassive black holes, but the cusps have not been detected 
outside of the Local Group since their angular sizes are less than 
$\sim 0.1''$.
We show that the presence of a cusp implies a lower limit of
$\sim 10^{-4}$ yr$^{-1}$ on the rate of stellar tidal disruptions,
and discuss the consequences of the cusps for
gravitational lensing and the distribution of dark matter on 
sub-parsec scales.

\end{abstract}

\section{Introduction}

Mass distributions near the centers of early-type galaxies 
are well described as power laws, $\rho\sim r^{-\gamma}$,
with indices $\gamma$ that change gradually with radius.
At their innermost resolved radii, most galaxies have
$0.5\lap\gamma\lap 2.$, with the steeper slopes
characteristic of fainter galaxies \citep{virgo-VI}.
If a supermassive black hole is present, the orbits of stars
will be strongly influenced at distances less than 
$\sim r_h= GM_\bullet/\sigma^2 \approx
10$ pc $(M_\bullet/10^8M_\odot)(\sigma/200 {\rm km\ s}^{-1})^{-2}$,
the black hole's gravitational influence radius.
Most galaxies are spatially unresolved on these small scales;
two clear exceptions are the nucleus of the Milky Way,
for which number counts extend inward to $\sim 0.002 r_h$
\citep{genzel-03}, and M32, which is resolved down to
a radius of $\sim 0.2 r_h$ \citep{lauer-98}.
Both galaxies exhibit steep density slopes,
$\gamma\approx 1.5$, at $r\lap r_h$.
Outside of the Local Group, only giant ellipticals
have sufficiently large black holes that $r_h$ can be
resolved; the nuclear luminosity profiles in
these galaxies are also power laws but very flat,
$\gamma \lap 1$.

Many distributions of stars are possible
around a black hole, but under two circumstances,
the stellar distribution at $r\lap r_h$ is predictable.
(1) If the black hole has been present for a time longer than
$T_r$, the relaxation time in the nucleus,
exchange of
energy between stars will drive the stellar distribution
toward a collisional steady state; assuming a single
stellar mass and ignoring physical collisions between stars,
this steady state has $\rho\sim r^{-7/4}$
at $r\lap r_h$ \citep{bw-76}.
(2) If the nucleus formed via the merger of two galaxies
each with its own supermassive black hole, the two black
holes will displace of order their combined mass in
the process of forming a tightly-bound pair \citep{mm-01},
producing a low-density core.
The first mechanism may be responsible for the steep
density profiles observed at the centers of the Milky Way and
M32, since both galaxies have central relaxation times
of order $10^9$ yr and both are near enough that
linear scales of order $r_h$ are well resolved.
The second mechanism may explain the very flat
central profiles of luminous E galaxies \citep{milos-02,ravin-02}; 
the central relaxation times of these galaxies are
much longer than $10^{10}$ yr and the stellar
distribution would be expected to remain
nearly unchanged after the two black holes had
coalesced into one.

In this paper we point out that both outcomes are possible.
A galaxy may form via
mergers, but at the same time, its central relaxation 
time following the merger may be shorter than $10^{10}$ yr.
In this circumstance, the cusp of stars
around the black hole is first destroyed
by the massive binary, then is regenerated via encounters
between stars in the gravitational field of the single,
coalesced hole. 
The result is a steep inner density profile in a galaxy
that had previously experienced the scouring effects
of a massive binary.
To the extent that {\it all} stellar spheroids experienced
mergers -- if only in the distant past -- this picture
is probably generic, applying even to small dense systems
like M32 and to the bulges of spiral galaxies like the Milky Way.
Understanding the conditions under which a previously-destroyed 
density cusp can spontaneously regenerate is crucial
if one wishes to interpret the present-day luminosity profiles of
galaxies as fossil relics of their merger histories \citep{volonteri-03}.

We use $N$-body simulations (\S2) to follow
first the destruction (\S3) and then the spontaneous regeneration (\S4)
of density cusps around black  holes.
The two most important free parameters in this problem
are the mass ratio $q\equiv M_2/M_1$ of the binary
black hole, and the slope $\gamma$ of the initial density 
cusp surrounding the larger hole.
We present results for several combinations of $q$ and
$\gamma$ (Table 1).
Our conclusion is that collisional, Bahcall-Wolf density
cusps should be ubiquitous in stellar spheroids
fainter than $M_V\approx -18.5$ that contain massive black holes,
essentially regardless of their merger histories.
However these cusps have gone undetected in galaxies outside the Local
Group because they are unresolved.
In \S 5 we discuss a number of consequences of the presence of the
cusps.

\begin{table*}
\begin{center}
\caption{Parameters of the $N$-body integrations \label{tbl-2}}
\begin{tabular}{ccccccc|ccccc}
\tableline\tableline
Run & $\gamma$ & $M_2/M_1$ & $r_{h_{1}}$ & $r_{h_{12}}$ & $T_r(r_{h_{12}})$ & $a_h$ & $r_h'$ & $T_r(r_h')$ & $T_r(0.2r_h')$ & $T_{gap}$ & $r_h''$\\
\tableline
1 & $0.5$ & $0.5$  & $0.264$  & $0.326$ & $1170.$  & $0.0181$  & $0.39$ & $1420.$ & $620$ & $79.$ & $0.38$ \\ 
2 & $0.5$ & $0.25$ & $0.264$  & $0.296$ & $1010.$  & $0.0119$  & $0.35$ & $1210.$ & $420$ & $48.$ & $0.34$ \\
3 & $0.5$ & $0.1$  & $0.264$  & $0.278$ & $916.$   & $0.00573$ & $0.32$ & $1070.$ & $390$ & $22.$ & $0.31$ \\
&&&&&&&&& \\
4 & $1.0$ & $0.5$  & $0.165$  & $0.210$ & $599.$   & $0.0116$  & $0.28$ & $870.$  & $300$ & $48.$ & $0.27$ \\
5 & $1.0$ & $0.25$ & $0.165$  & $0.188$ & $499.$   & $0.00751$ & $0.23$ & $640.$  & $270$ & $26.$ & $0.23$ \\
6 & $1.0$ & $0.1$  & $0.165$  & $0.174$ & $441.$   & $0.00360$ & $0.20$ & $520.$  & $220$ & $11.$ & $0.20$ \\
&&&&&&&&& \\
7 & $1.5$ & $0.5$  & $0.0795$ & $0.107$ & $217.$   & $0.00594$ & $0.17$ & $420.$  & $160$ & $23.$ & $0.17$ \\
8 & $1.5$ & $0.25$ & $0.0795$ & $0.093$ & $170.$  & $0.00364$ & $0.13$ & $260.$  & $110$ & $10.$ & $0.13$ \\
9 & $1.5$ & $0.1$  & $0.0795$ & $0.085$ & $144.$  & $0.00176$ & $0.11$ & $200.$  & $ 75$ & $4.1$ & $0.11$ \\
\tableline
\end{tabular}
\end{center}
\end{table*}

\section{Models and Methods}

We started by constructing Monte-Carlo realizations of
steady-state galaxy models having Dehnen's (1993) density law, 
with an additional, central point mass representing a black
hole.
The Dehnen-model density follows $\rho(r) \propto r^{-\gamma}$
at small radii, and the isotropic phase-space distribution function
 that reproduces Dehnen's $\rho(r)$ 
in the presence of a central point mass is non-negative for all 
$\gamma\ge 0.5$;
hence $\gamma=0.5$ is the flattest central profile that can be adopted
if the initial conditions are to represent an isotropic, steady state.
We considered initial models with $\gamma=(0.5,1.0,1.5)$.
The mass $M_1$ of the central ``black hole'' was always $0.01$,
in units where the total mass in stars $M_{gal}$ was one;
the Dehnen scale length $r_D$ and the gravitational constant 
$G$ were also unity.
The $N$-body models so constructed were in a precise
steady state at time zero.

Destruction of the cusp was achieved by introducing a second
``black hole'' into this model, which spiralled into the center,
forming a binary with  the first (more massive) hole
and displacing stars.
Three values were used for the mass of the smaller hole:
$M_2/M_1\equiv q =(0.5,0.25,0.1)$.
The smaller hole was placed initially at a distance
$1.6$ from the center, with a velocity roughly
$1/2$ times the circular velocity at that radius;
a non-circular orbit was chosen in  order to speed up
the orbital decay.

After the orbit of the smaller black hole had decayed via
dynamical friction against the stars, it formed a tight
binary with the more massive hole, with a relative orbit
close to circular.
An estimate of the semi-major axis $a_h$ at which the
binary first becomes ``hard'' is $a_h = G \mu/4\sigma^2$
where $\mu\equiv M_1M_1/(M_1+M_2)$ is the reduced mass.
The precise meaning of ``hard'' is debatable; 
the definition just given defines
a ``hard'' binary as one whose binding energy per unit
mass, $|E|/(M_1+M_2)$, exceeds $2\sigma^2$.
While simple, this definition contains the ill-defined quantity
$\sigma$, which is a steep function of position near the black hole(s).
We followed \cite{wang-05} and used the alternative definition
\begin{equation}
a_h \equiv {\mu\over M_1+M_2}{r_{h_{12}}\over 4} = {q\over (1+q)^2}{r_{h_{12}}\over 4},
\label{eq:ah}
\end{equation}
with $r_{h_{12}}$ the gravitational influence radius defined below.
In practice, $a_h$ so defined was found to be roughly 
(within a factor $\sim 2$) the value of the semi-major axis at 
which the binary hardening rate $(d/dt)(1/a)$ first became 
approximately constant.

Decay was allowed to continue until the binary semi-major axis
had reached a value of $a_h/5\approx qr_h/20$.
At this point, the two black holes were replaced by a single 
particle of mass $M_{12}=M_1+M_2$,
with position and velocity given by the center of mass of the binary.
The $N$-body integration was then continued for a time roughly
equal to the relaxation time $T_r(r_h')$ defined below.
The most suitable time at which to merge the two black holes
was not clear {\it a priori}; our choice is of order the separation at which
gravitational-wave emission would induce coalescence in
$\sim 10^{10}$ yr \citep{living-05}, but in fact we expect
that other processes like interaction of the binary with
ambient gas may drive the final coalescence in real galaxies,
and it is not clear at what separation these processes are
likely to dominate the evolution.

The ratio $(M_1+M_2)/M_{gal}\approx 0.01$ in our models is roughly a factor
ten larger than the ratio of black hole mass to galaxy mass
in real spheroids \citep{mf-01a,marconi-03}.
This is acceptable as long as we are careful to present masses and radii
in units scaled to $M_1+M_2$ when making comparisons with real
galaxies.

Table 1 gives a number of parameters associated with
the $N$-body integrations.
The gravitational influence radius $r_h$ was defined
as the radius containing a mass in stars equal to 
twice the mass $M_\bullet$ of the central black hole.
This definition, while superior to $GM_\bullet/\sigma^2$, 
is somewhat ambiguous in our $N$-body models,
given that the effective mass of the central object, 
and the distribution of the stars, both change with time.
We accordingly defined four different influence radii.
(1) At the start of the integrations, the larger black
hole, of mass $M_1$, was located at the center of the
galaxy. 
Its influence radius $r_{h_{1}}$ was computed by setting 
$M_\bullet = M_1$ and using the $t=0$ stellar distribution.
(2) After the smaller black hole has fallen in to
a distance $\lap r_{h_{1}}$ from the larger hole,
the appropriate value of $M_\bullet$ becomes $M_1+M_2$.
We defined the associated influence radius to be $r_{h_{12}}$,
which we computed ignoring the changes that had occurred
in the stellar distribution since $t=0$.
(3) The third influence radius, $r_h'$, was
computed by setting $M_\bullet=M_1+M_2$, but this mass was
compared with the stellar distribution at the end of the binary
evolution phase, after the phase of cusp destruction.
(4) Finally, $r_h''$ is the influence radius at
the end of the second phase of integration, after 
a Bahcall-Wolf cusp has formed around the single
black hole.
As Table 1 shows, $r_h''$ is only slightly smaller than
$r_h'$ since the regenerated cusp contains a mass that
is small compared with $M_\bullet$.

\begin{figure*}
\centering
\includegraphics[scale=0.70,angle=-90.]{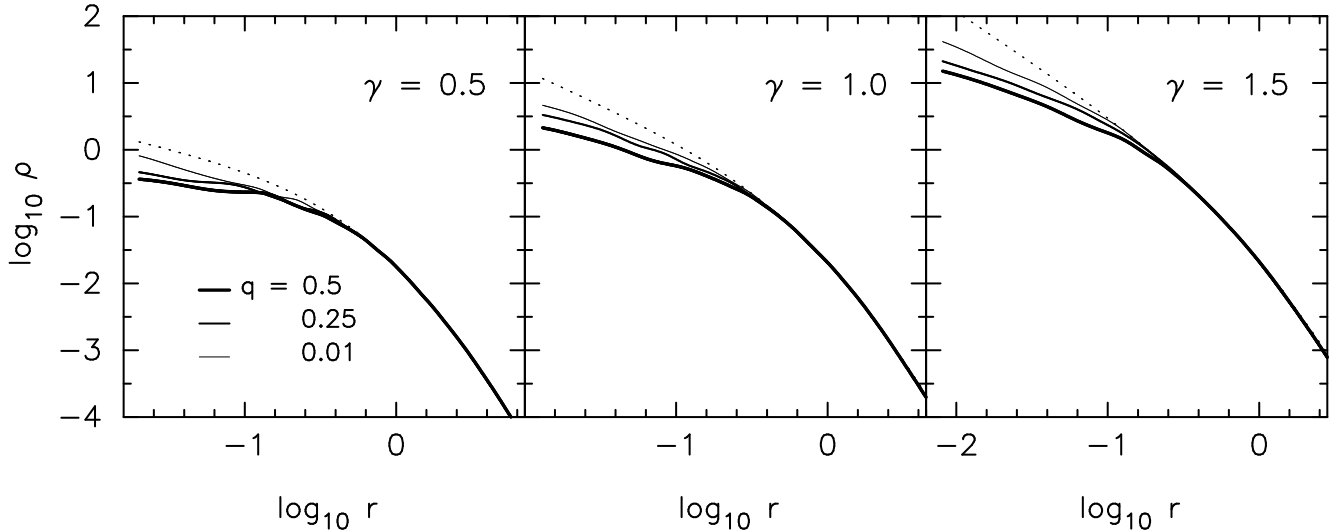}
\caption{Cusp destruction.
Solid lines are stellar density profiles just before the two
``black holes'' were combined into one.
Dashed lines show the initial models.
$q=M_2/M_1$ is the binary mass ratio.
\label{fig:profs}
}
\end{figure*}

The relaxation times $T_r$ in Table 1 were computed
from the standard expression (eq. 2-62 of Spitzer 1987), 
setting $\ln\Lambda = \ln(\sigma^2r_{h12}/2Gm_\star)$,
with $\sigma$ the 1D stellar velocity dispersion at
$r_{h12}$ and $m_\star=N^{-1}$ the mass of an $N$-body particle.
This definition of $\Lambda$ is equivalent to equating
$b_{max}$, the maximum impact parameter for encounters
in Chandrasekhar's theory, with $r_h$ \citep{preto-04}.
Table 1 gives values of $T_r$ evaluated at two of the
four influence radii defined above. 
$T(r_{h_{12}})$ was computed using the structural parameters
of the initial galaxy model,
while $T_r(r_h')$ was computed using estimates 
like those in Figures~\ref{fig:profs} and~\ref{fig:kinem}.
of the stellar density and velocity dispersion in the evolved models.
We also give $T_r(0.2rh')$; the motivation for this is given
below.
The final time scale in Table 1, $T_{gap}$, is
defined below (equation {\ref{eq:tgap}) and is an
estimate of the time required for the angular-momentum
gap created by the binary to be refilled;
$T_{gap}$ varied between $\sim T_r(r_h)/20$ 
and $\sim T_r(r_h)/50$.

The power-law cusps in our initial models
were motivated by the approximately
power-law dependence of luminosity density on radius
observed near the centers of many early-type galaxies 
\citep{virgo-VI}.
Since the observations often do not resolve $r_h$,
the stellar distribution at $r\lap r_h$ in some galaxies
might be different than the inward extrapolation of the
power laws that are fit to larger radii.
For instance, galaxies with sufficiently short relaxation
times are expected to have $\rho\sim r^{-7/4}$
density cusps like the ones that form in our $N$-body
models at late times.
Other galaxies may have compact stellar nuclei \citep{virgo-VIII}.
We did not include such dense features in our initial models:
first, because the associated mass would have been small 
compared with the mass removed by the binary; and second,
because doing so would have more than doubled the computational
effort due to the short time steps required for stars initially
near the black holes.

All $N$-body integrations used $N=0.12\times 10^6$ particles
and were carried out on a GRAPE-6 special-purpose computer.
The $N$-body integrator is described in \cite{msm-06}.
This algorithm is an adaptation of {\tt NBODY1}
\cite{Aarseth:99} to the GRAPE-6; it uses a fourth-order 
Hermite integration scheme with individual, adaptive, 
block time steps \citep{Aarseth:03}.
For the majority of the particles, the forces and force
derivatives were calculated via a direct-summation scheme on
the GRAPE-6, using the particle advancement scheme described
in \cite{bms-05} with an accuracy parameter of $\eta=0.01$
and zero softening.
Close encounters between the black holes, 
and between black holes and stars,
require prohibitively small time steps in such a scheme
and were regularized using the chain regularization
routine of Mikkola and Aarseth \citep{MA:90,MA:93}.
A detailed description of the chain algorithm, including
the results of performance tests, are given in 
\cite{msm-06}.

\begin{figure*}
\centering
\includegraphics[scale=0.7,angle=-90.]{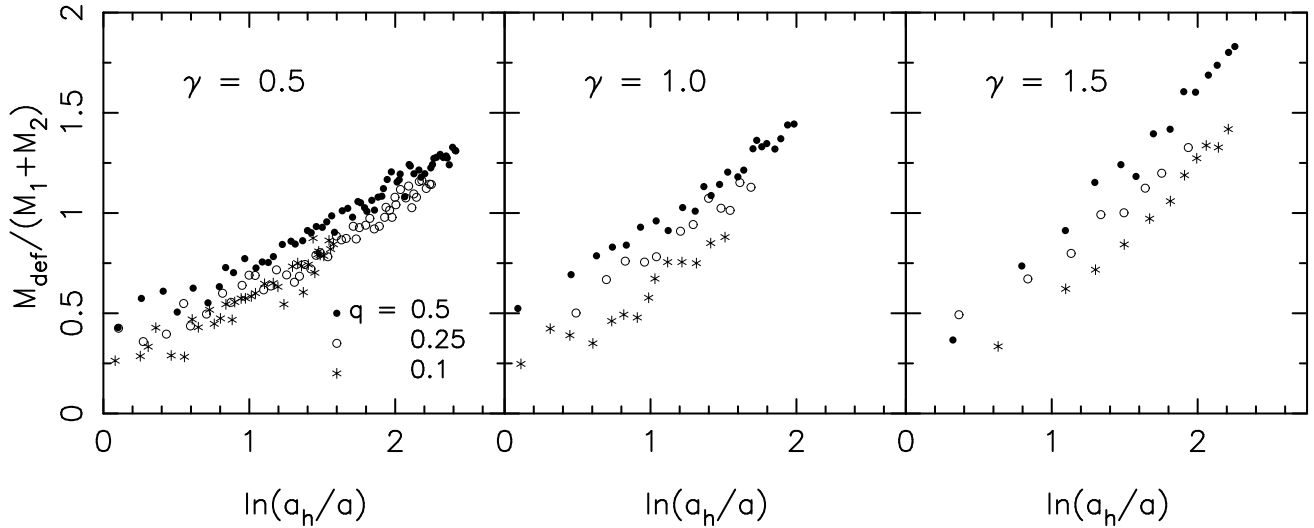}
\caption{Mass deficits, computed as described in the text, 
during the binary phase of the integrations; $a=a(t)$ is the
binary semi-major axis and $a_h$ is defined in equation (\ref{eq:ah}).
$q=M_2/M_1$ is the binary mass ratio.
\label{fig:mdef}
}
\end{figure*}

Our initial conditions (one black hole at the center,
a smaller black hole orbiting about it) are not
as realistic as in simulations that follow both
merging galaxies from the start 
(e.g. \cite{mm-01}, \cite{merritt-02}),
but are superior to simulations that
drop one or two black holes into a pre-existing galaxy
that contains no black hole
(e.g. \cite{quinlan-97,nakano-99a,nakano-99b}).
We ignore the radiation recoil that would accompany
the final coalescence of the two black holes,
displacing the remnant hole temporarily from its central
location and increasing the size of the core 
\citep{merritt-04b,boylan-04}.
We also ignore processes like loss of stars into the
black hole(s), stellar tidal disruptions, and
stellar collsions, all of which might affect the
form of the final density profile.

\begin{figure*}
\centering
\includegraphics[scale=0.7,angle=-90.]{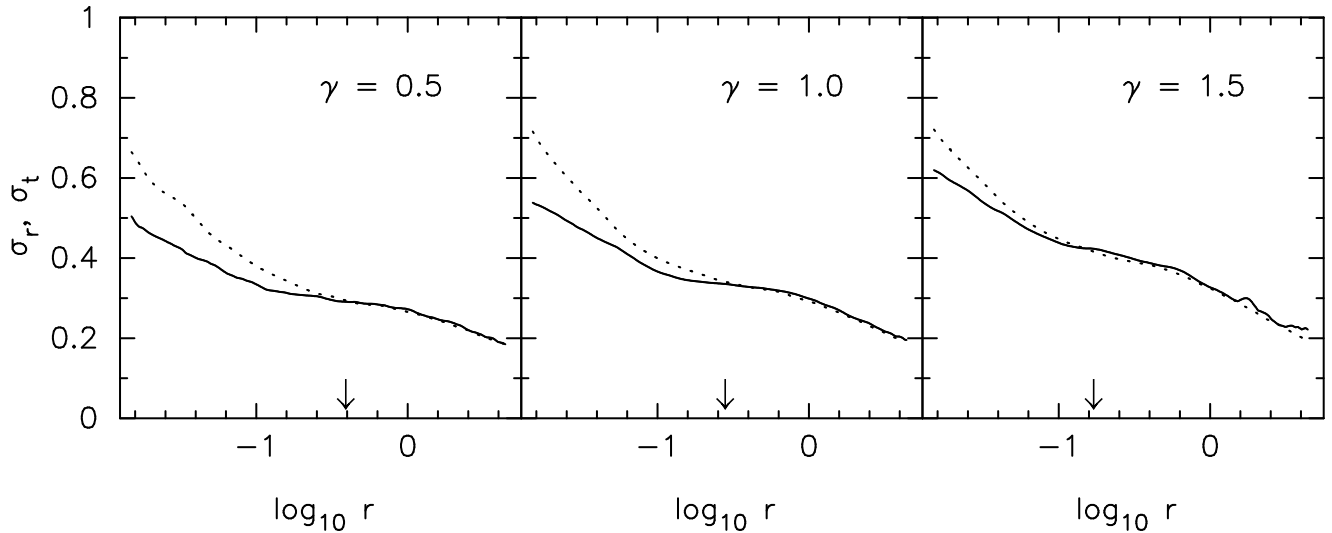}
\caption{Stellar velocity dispersions at the end of
the binary phase, in the integrations with $M_2/M_1=0.5$.
Solid lines: $\sigma_r$; dashed lines: $\sigma_t$.
The binary creates a tangentially-biased velocity 
distribution near the center by preferentially 
ejecting stars on radial orbits.
Arrows indicate the black hole influence radius $r_h'$.
\label{fig:kinem}
}
\end{figure*}

\section{Cusp Destruction}

After formation of a hard binary at $a\approx a_h$,
the binary's binding energy continues to
increase as the two massive particles eject stars
via the gravitational slingshot \citep{saslaw-74,mv-92,quinlan-96}.
As in a number of recent studies \citep{bms-05,msm-06},
we found that the hardening rates were nearly independent of
 time, $s\equiv (d/dt)(1/a) \approx $ const., for $a<a_h$.
The values of $s$ are not of particular interest here, since
$N$-body hardening rates are known to depend strongly on 
particle number\citep{bms-05}, and we do not expect
these simulations to be in the ``empty loss cone'' regime
that characterizes binary evolution in real
galaxies \citep{mm-03}.
However the ``damage'' done by the binary to the pre-existing
stellar distribution is expected on energetic grounds
to be a function of $a_h/a$, essentially independent of the {\it rate} of
hardening and hence of $N$ \citep{quinlan-96}, and
this expectation has recently been confirmed in $N$-body
experiments \citep{msm-06}.
As described above, the binary phase of the integrations
was terminated when $a=a_h/5$.

Figure~\ref{fig:profs} shows density profiles of the $N$-body
models at the end of the binary phase.
Radii of the ``star'' particles 
were computed relative to the center of mass of the binary,
and estimates of $\rho(r)$ were constructed using
the  nonparametric kernel estimator of \cite{mt-94}.
The damage done by the infalling black hole can be seen
to increase with its mass for $\gamma=0.5$ and $1$, 
and extends out to a radius $\sim r_h$.
The $\gamma=0.5$ cusp is converted into an approximately
constant-density core, while the $\gamma=1$ and $\gamma=1.5$ 
cusps are ``softened,'' to power laws of index $\sim 0.5$
($\gamma=1$) and $\sim 1.0$ ($\gamma=1.5$).

A standard measure of the damage done by an infalling
black hole to a pre-exising density cusp is the ``mass
deficit'' $M_{def}$, defined as the decrease in the central 
mass within a sphere that contains the affected region \citep{milos-02}.
The mass deficit is potentially observable 
\citep{milos-02,ravin-02,graham-04}, 
assuming that one can guess the pre-existing density profile,
and its value is an index of the cumulative effect of mergers 
on the galaxy \citep{volonteri-03}.
Since there are few results in the literature on
the sizes of mass deficits generated by large-mass-ratio inspirals,
we show in Figure~\ref{fig:mdef} mass deficits
for these integrations, expressed in terms of $a_h/a(t)$;
we extended some of the integrations beyond $a=a_h/5$ in
order to further elucidate this dependence.
Figure~\ref{fig:mdef} shows that -- for a given degree of binary 
hardness --
the mass deficit is much better predicted by $M_1$ or $M_{12}$
than by $M_2$.
In other words, the damage done by the smaller black hole
is rougly proportional to the mass of the {\it binary},
at a given value of $a_h/a$.

\begin{figure*}
\centering
\includegraphics[scale=0.7,angle=-90.]{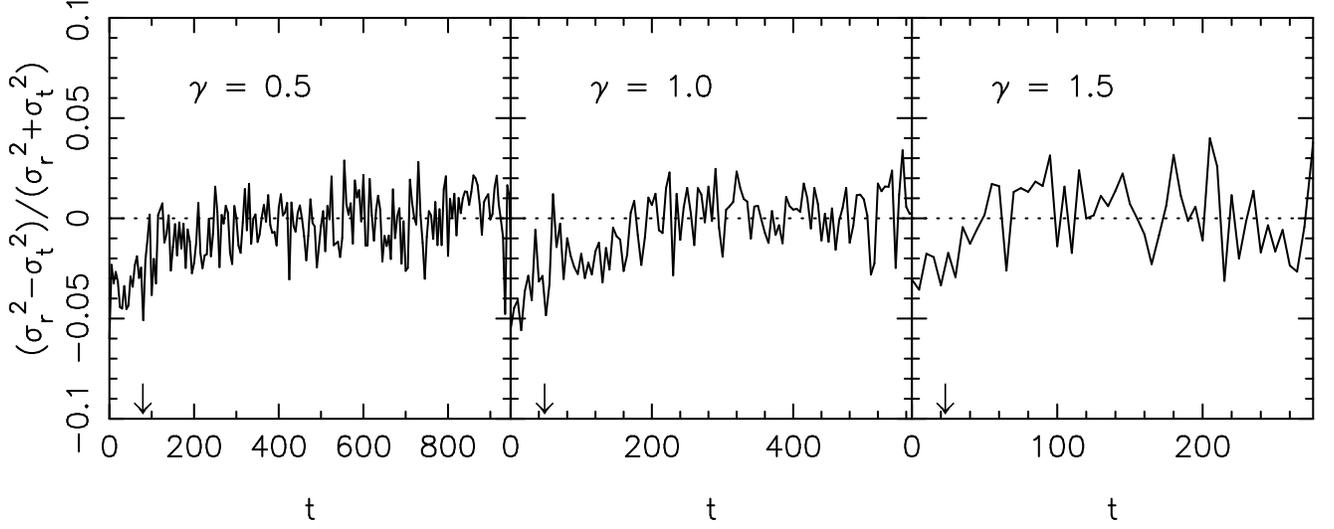}
\caption{Mean velocity anisotropy of the stars in a sphere
of radius $r_h'$ around the single black hole, as
a function of time after the massive binary was combined
into a single particle.
Shown are the integrations with $M_2/M_1=0.5$.
Arrows indicate the time $T_{gap}$ defined in the text
(equation \ref{eq:tgap}, Table 1).
\label{fig:anisot}
}
\end{figure*}

\begin{figure*}
\centering
\includegraphics[scale=0.85]{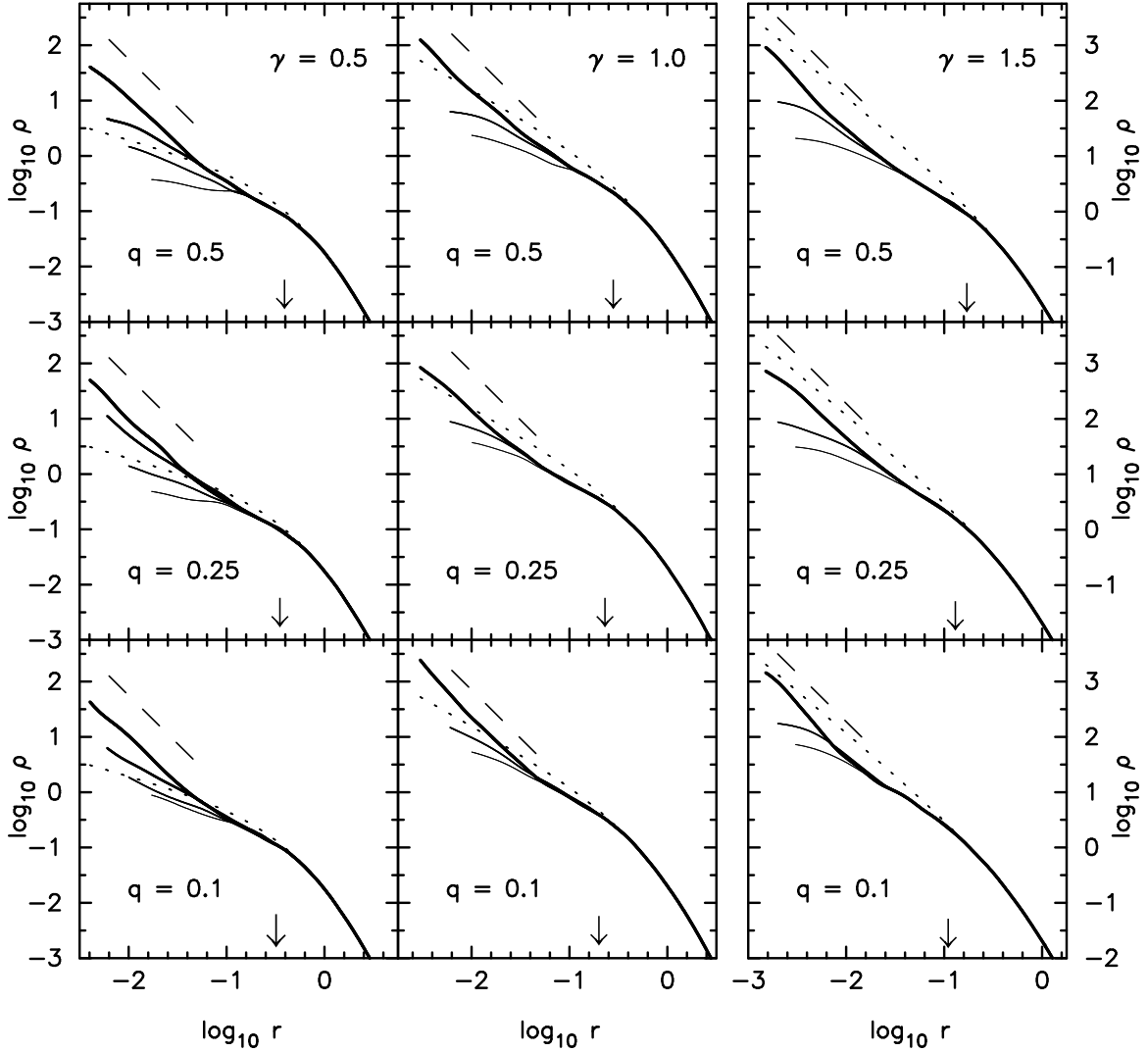}
\caption{
Cusp regeneration.
Plots show the stellar density profile around the
single black hole, at post-coalescence times
of approximately $(0,0.1,0.5,1)T_r(r_h)$ ($\gamma=0.5$),
$(0,0.2,1)T_r(r_h)$ ($\gamma=1.0$),
and $(0,0.2,0.5)T_r(r_h)$ ($\gamma=1.5$).
Dotted lines show the density profiles at the start
of the binary integrations, when only one black
hole was located at the center.
Dashed lines have logarithmic slope of $-1.75$.
Arrows indicate $r_h'$, the influence radius of
the single black hole just after coalescence of the
binary (Table 1).
\label{fig:grow}
}
\end{figure*}

\begin{figure}
\centering
\includegraphics[scale=0.5]{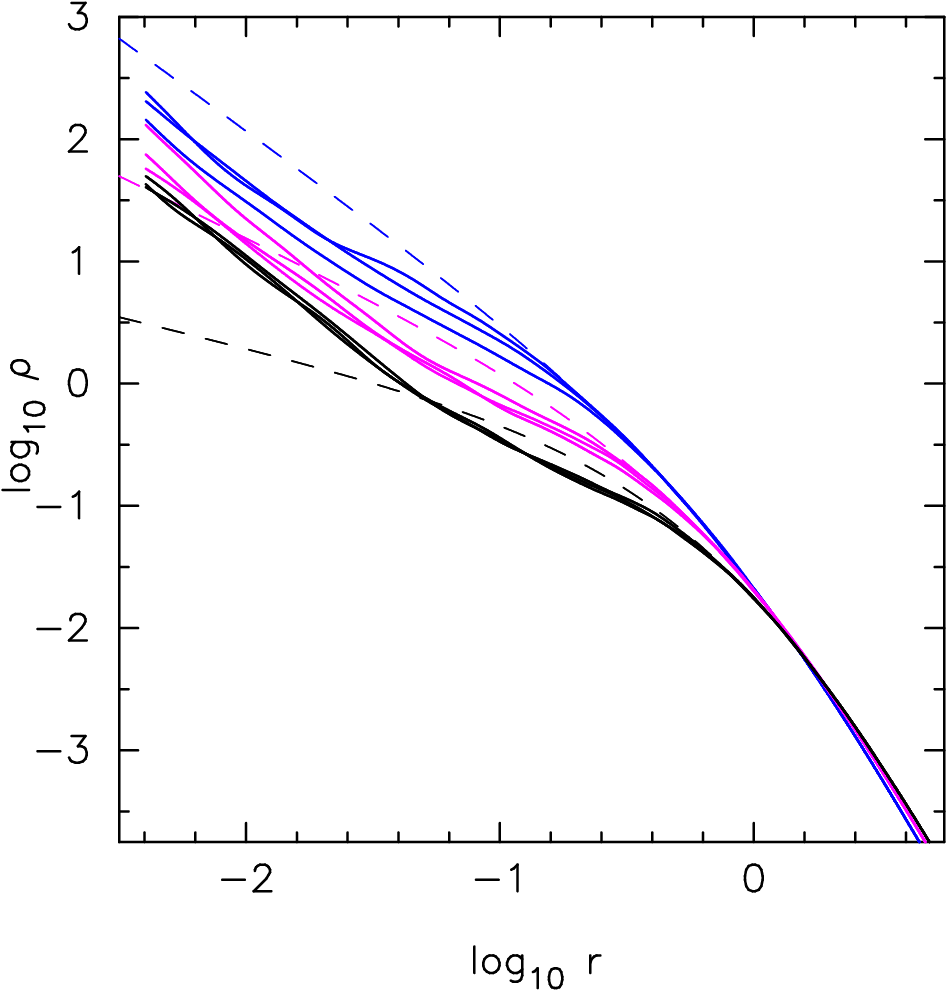}
\caption{Final density profiles after
growth of the cusp.
Black lines had $\gamma=0.5$ for the pre-binary
initial conditions, red lines had $\gamma=1.0$,
and blue line had $\gamma=1.5$.
Dashed lines show the initial models.
\label{fig:all}
}
\end{figure}

This is a reasonable result.
Equating the increase in the binary's binding energy with
the energy lost to an ejected star gives
\beq
{G\mu(M_1+M_2)\over 2} \delta\left({1\over a}\right) =
(E_f-E_i)\delta M_{ej}
\eeq
with $E_i$ and $E_f$ the initial and final kinetic 
energy per unit mass of the ejected star and $\delta M_{ej}$ its mass;
$\mu$ is the reduced mass of the binary.
This can be written
\beq
{dM_{ej}\over d\ln(1/a)} = {\mu\over 2} {V_{bin}^2\over E_f-E_i}
\eeq
where $V_{bin}=\sqrt{G(M_1+M_2)/a}$  is
the relative velocity between the components of the
binary (assuming a circular orbit).
Hills (1983) finds from scattering experiments that, 
for a hard binary in the limit $M_2\ll M_1$,
$E_f-E_i$ for a zero-impact-parameter encounter is $\sim 1.4 G\mu/a$.
This implies
\beq
{dM_{ej}\over d\ln(1/a)} \approx 0.36 (M_1+M_2)
\label{eq:hills1}
\eeq
and the total mass ejected by the binary in hardening 
from $a\approx a_h$ to $a_f$ is
\beq
{\Delta M_{ej}\over M_1+M_2} \approx 0.36 \ln \left({a_h\over a_f}\right),
\label{eq:hills2}
\eeq
i.e., the mass ejected in reaching a given degree of hardening $a_h/a$
is proportional to $M_1+M_2$.
Quinlan (1996) used scattering experiments to compute $J$
in the relation
\beq
{dM_{ej}\over d\ln(1/a)} = J(M_1+M_2)
\label{eq:mejrate}
\eeq
as a function of $M_2/M_1$, assuming a Maxwellian
distribution of velocities at infinity.
He defined ``ejected'' stars to be those with
final velocities exceeding $\max\{1.5V_0,\sqrt{3}\sigma\}$;
here ``final'' means after the star has moved
far from the binary (ignoring any velocity changes that
result from the galactic potential), $V_0$ is
the velocity before the encounter, and $\sigma$ is the
velocity dispersion far from the massive binary.
Quinlan's definition is a reasonable one if $M_{ej}$ is to be
equated with $M_{def}$, and Quinlan found that $J$ is nearly independent 
of $M_2/M_1$ for a hard binary, decreasing from $J\approx 1$
for $M_2=M_1$ to $J\approx 0.5$ for $M_2=M_1/256$ (his Figure 5).
These $J$-values are consistent with Figure~\ref{fig:mdef}.

Expressions like (\ref{eq:hills2}) seem counter-intuitive,
since they seem to imply an ejected mass that is large even for
small $M_2$.
However for small $M_2$, $a_h\propto M_2/M_1$ (equation \ref{eq:ah}), 
so for a given $a_f$, the ejected mass is predicted
to scale as $\sim\ln M_2$.
The dependence of $M_{def}$ on binary mass ratio is a
topic that deserves further study since the central structure
of bright elliptical galaxies is expected to reflect the cumulative
effect of many minor mergers
\citep{volonteri-03}.

We do observe an additional dependence of $M_{def}$
on both $\gamma$ and $q$ at given $a_h/a$ (Figure~\ref{fig:mdef}).
However the definition of $a_h$ is somewhat arbitrary
(\S2) and we do not pursue that additional dependence here.

The binary preferentially ejects stars on eccentric orbits.
The result is an induced anisotropy in the stellar
velocity distribution at $r\lap r_h$.
Figure~\ref{fig:kinem} illustrates this, for the 
integrations with the most massive secondaries
($M_2/M_1=0.5$).
The induced anisotropy is mild, similar to what has
been observed in other $N$-body studies
(e.g.\cite{mm-01}).

\section{Cusp Regeneration}

When the separation between the two components of the
binary had dropped to $a=a_h/5$,
the two massive particles were combined into one
and placed at the center of mass of the binary.
The integrations were then continued until an
additional time of roughly $T_r(r_h')$ had elapsed
(Table 1).
During this second phase of the integrations, the
stellar distribution around the single black hole
evolved due to star-star encounters.
The initial effect of encounters was to refill
the phase space gap created when the binary
ejected stars with pericenters $r_p\lap a_h$.
The refilling time is approximately
\beq
T_{gap}\approx {a_h\over r_h} T_r(r_h) \approx {1\over 4} 
{q\over (1+q)^2} T_r(r_h)
\label{eq:tgap}
\eeq 
\citep{wang-05}.
Table 1 shows that $T_{gap}$ varies between $\sim 10$ ($\gamma=1.0, q=0.1$)
and $\sim 80$ ($\gamma=0.5,q=0.5$) in our units,
or between $\sim 0.02T_r(r_h)$ and $\sim 0.05T_r(r_h)$.
Figure~\ref{fig:anisot} verifies that the initially anisotropic
velocity distribution at $r\lap r_h$ becomes isotropic
in a time of order $T_{gap}$.
In luminous galaxies, even $T_{gap}$ can exceed
a galaxy lifetime \citep{wang-05}, and such galaxies
will experience a greatly reduced rate of interaction
of stars with the black hole compared to galaxies
with a full loss cone.

On the longer time scale of $\sim T_r(r_h)$, energy
exchange between stars modifies their radial distribution,
eventually reaching the zero-flux Bahcall-Wolf (1976) solution
$\rho\sim r^{-7/4}$.
This is shown in Figure~\ref{fig:grow} for the 
$\gamma=0.4$ and $\gamma=1$ models.
(Results for $\gamma=1.5$ are shown in Figures~\ref{fig:all} 
and~\ref{fig:MW} and are discussed separately below.)
The new cusp attains the $\rho\sim r^{-7/4}$ form inside
of $\sim 0.2 r_h'$.
In all of the integrations, the logarithmic
slope of the final cusp is steeper than that
of the initial (pre-binary) galaxy at radii $r\lap 0.5 r_h'$,
although the final density remains below the initial
density at $r\approx r_h'$.

Figure~\ref{fig:all} shows the final density profiles
for all seven models.
There is remarkably little dependence of the final
result on the mass ratio of the binary.
The reason can be seen by comparing Figures~\ref{fig:profs}
and~\ref{fig:all}: while the different mass ratios
do yield different density profiles at the end of the
binary phase, most of these differences are at $r\lap 0.3 r_h$,
and this region is efficiently refilled by the new cusp.

The final density profiles in Figure~\ref{fig:all}
can be well approximated as broken power laws 
at radii $r\lap r_h$:
\begin{eqnarray}
\rho(r) &=& \rho(r_0)\left({r\over r_0}\right)^{-7/4},\ \ \ \ r\le r_0, \nonumber\\
&=& \rho(r_0)\left({r\over r_0}\right)^{-1},\ \ \ \ r_0<r\lap r_h.
\label{eq:spike}
\end{eqnarray}
The ``break'' radius $r_0$ is $\sim 0.15 r_h''$ in the models
with initial cusp slope $\gamma=0.5$, $\sim 0.20 r_h''$
for $\gamma=1.0$, and $\sim 0.25r_h''$ for $\gamma=1.5$.
In this approximation to $\rho(r)$, the mass in the cusp,
$M_{cusp}\equiv M(r\le r_0)$, is given by
\beq
M_{cusp} = {2M_\bullet\over 1 + 0.75\left[\left(r_h/ r_0\right)^2-1\right]}.
\eeq
Thus $M_{cusp}/M_\bullet\approx 0.03$ ($\gamma=0.5$),
$\sim 0.12$ ($\gamma=1.0$), and $\sim 0.16$ ($\gamma=1.5$), 
confirming that $M_{cusp}$ is small compared with $M_\bullet$.
One consequence is that $r_h''$, the influence radius at the final
time step, differs only slightly from $r_h'$, the influence
radius at the end of the binary phase (Table 1).

Finally, we consider how much time is required to regrow the cusps.
As noted above, all of the post-binary integrations were
carried out until an elapsed time of $\sim T_r(r_h')$,
and this was also roughly the time required for the cusp
to reach its steady-state form.
However the cusp extends only to a radius of $\sim 0.2r_h'$,
and a more relevant estimate of the relaxation time is
probably $T_r(0.2r_h')$.
Estimates of this time are given in Table 1.
The time to fully grow the cusp is approximately
$2-3$ times the relaxation time at $0.2 r_h'$.
As Figure~\ref{fig:grow} shows, considerable regeneration
takes place even in a much shorter time, hence we might
predict the presence of steep central profiles even in galaxies
substantially younger than either $T_r(r_h)$ or $(2-3)T_r(0.2r_h)$.
We return to this issue below.

\section{Implications}

\begin{figure*}
\centering
\includegraphics[scale=0.7,angle=-90.]{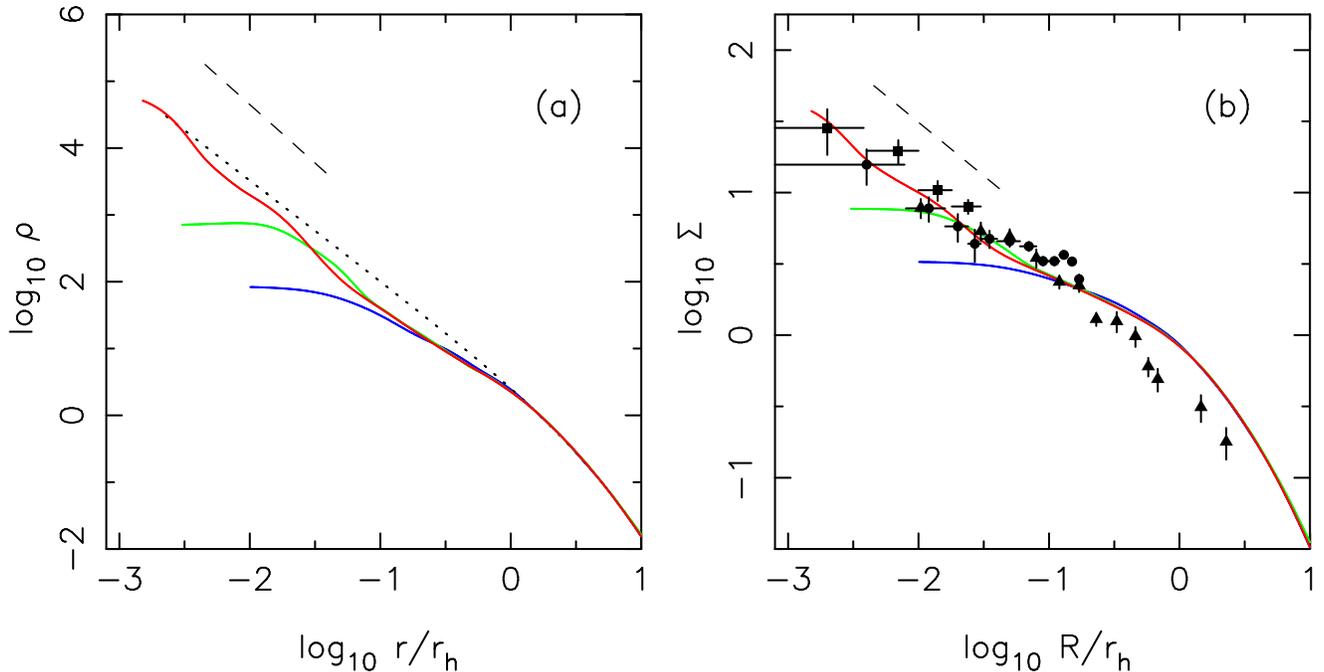}
\caption{
Evolution of the spatial (a) and projected
(b) density profiles in the simulation that most closely
resembles the Milky Way nucleus.
The initial model (dotted line) had a $\rho\sim r^{-1.5}$
density cusp and a central black hole of mass $M_1=0.01$,
in units where the total stellar mass of the model is 1.
Blue (lower) line shows the density after infall and capture 
of a second black hole of mass $M_2=0.1M_1$.
Green and red lines show the evolving density
at times ($40,100$) after coalesence; 
based on the discussion in the text, the corresponding
physical times are roughly ($3,8$) $\times 10^9$ yr.
Symbols in (b) show the observed surface density of stars
near the Galactic Center, from Genzel et al. (2003);
see Figure 7 from that paper for symbol definitions.
The vertical normalization of the symbols was set
by eye in order to give a good match to the $N$-body
data at $R\lap 0.2r_h$.
Dashed lines have logarithmic slopes of $-1.75$ (a)
and $-0.75$ (b).
\label{fig:MW}
}
\end{figure*}

\subsection{Cusp Regeneration in Local Group Galaxies}

The relaxation time in the Milky Way nucleus is less
than $10^{10}$ yr
and a number of authors have suggested that the stellar
cluster around the supermassive black hole 
might be collisionally relaxed
\citep{alexander-99,genzel-03}.
The stellar mass density, based
on number counts that extend down to $\sim 0.005$ pc, is
\beq
\rho(r) \approx 1.2\times 10^6 M_\odot {\rm pc}^{-3}
\left({r\over 0.38{\rm pc}}\right)^{-\alpha}
\label{eq:rho_MW}
\eeq
\citep{genzel-03}, 
with  $\alpha\approx 2.0$ at $r\ge 0.38$ pc 
and $\alpha\approx 1.4$ at $r<0.38$ pc.
The black hole mass is $3.7\pm 0.2\times 10^6M_\odot$
\citep{ghez-05}. 
The implied influence radius, defined
as the radius containing a mass in stars that is
twice the black hole mass, is 
$r_h\approx 88''\approx 3.4$ pc.

We used equation (\ref{eq:rho_MW}) and the measured value
of $M_\bullet$ to compute the stellar velocity dispersion
profile in the Milky Way nucleus, assuming a spherical
velocity ellipsoid.
The relaxation time $T_r$ was then calculated as in the $N$-body models, 
i.e.
\begin{subequations}
\begin{eqnarray}
&&T_r(r) = {0.338\sigma(r)^3\over \rho(r) m_\star G^2\ln\Lambda} \\
&&\approx 1.69\times 10^9 {\rm yr} \times \\ 
&& \left({\sigma(r)\over 100\ {\rm km\ s}^{-1}}\right)^3
\left({\rho(r)\over 10^6M_\odot}\right)^{-1}
\left({m_\star\over 0.7M_\odot}\right)^{-1}
\left({\ln\Lambda\over 15}\right)^{-1}
\end{eqnarray}
\end{subequations}
\label{eq:tr}
with $m_\star=0.7M_\odot$ and 
$\Lambda = \sigma^2(r_h)r_h/2Gm_\star$.
The relaxation time 
at $r=r_h$ was found to be $5.4\times 10^{10}$ yr,
dropping to $\sim 6\times 10^9$ yr at $0.2r_h$ and
$\sim 3.5\times 10^9$ yr at $\sim 0.1r_h$;
at smaller radii, $T_r$ increases very slowly with decreasing $r$.

As noted above, cusp regeneration in the $N$-body models required a time
of $\sim 1 T_r(r_h)$, or $\sim 2-3 T_r(0.2r_h)$.
However these two timescales are substantially different in
the Milky Way, suggesting that the structure of the Milky
Way nucleus differs in some respect from that of the
$N$-body models in the region $0.2r_h\lap r\lap r_h$.

Figure~\ref{fig:MW} makes the comparison.
We show the evolution of $\rho(r)$
and $\Sigma(R)$ (the latter is the projected density)
in the post-binary integration with parameters
($\gamma=1.5, q=0.5$);
the right-hand frame also shows the measured surface density
of stars at the Galactic center, 
from the number counts of \cite{genzel-03},
with arbitrary vertical normalization.
The $N$-body model successfully reproduces the slope of
the observed stellar cusp at late times,
but the $N$-body profile is less steep than the
observed counts at $r\gap 0.2r_h$, i.e., outside of the cusp.
In the Milky Way, an approximately power-law dependence
of $\rho$ on $r$ extends well beyond $r_h$, while
the $N$-body models exhibit a shallower slope at these radii.
Better correspondence between $N$-body model and data could presumably 
have been achieved by modifying the initial model, or by repeating the
integrations with smaller $M_\bullet$ (which would have
mandated a larger $N$ in order to resolve the smaller cusp); 
the unphysically large value of $M_\bullet$ adopted here
($M_\bullet=0.015M_{gal}$)
implies an influence radius that is comparable to the
Dehnen-model scale length.

In light of these ambiguities, it is reasonable to
fix the physical unit of time by comparing
the $N$-body model to the Milky Way at 
$\sim 0.2 r_h$; this is roughly the outer radius of
the cusp, and Figure~\ref{fig:MW} shows
that the model fits the data well out to this radius.
We find that $T_r(0.2r_h) \approx 160$ in $N$-body units,
while in the Milky Way, $T_r(0.2r_h)\approx 6\times 10^9$ yr.
The integration time of 300 in Figure~\ref{fig:MW} then
corresponds to a physical time of $\sim 1.1\times 10^{10}$ yr.
In fact, the $N$-body cusp attains nearly its final form by
a time of $\sim 200$, or $\sim 8\times 10^9$ yr.
We conclude that the distribution of stars near the Milky
Way black hole is consistent with a merger
having occurred at a time $\gap 8$ Gyr in the past.

In fact, the probability is $\sim 68\%$ that a galaxy with 
a dark-matter halo as large as that of the Milky Way has 
experienced a major merger (mass ratio of at least $4:1$) 
since a redshift of $z=2$, i.e., in the last $\sim 10$ Gyr 
\citep{merritt-02}.
If this occurred, our results suggest the interesting
possibility that the Milky Way
cusp may still be evolving toward its steady-state form.
Evidence for this might be sought in the form of a flattening
the density profile at very small radii, $r\lap 0.1r_h\approx 10''$.

The stellar density near the center of M32 is
\beq
\rho(r) \approx 2.2\times 10^5 M_\odot {\rm pc}^{-3}
\left({r\over 1{\rm pc}}\right)^{-\alpha},  \ \ \ \alpha\approx 1.5
\eeq
\citep{lauer-98}, very similar to that of the Milky Way.
Unfortunately the estimation of $M_\bullet$ in M32 suffers
from the degeneracy inherent in orbital modelling of
axisymmetric systems and the black hole mass
could lie anywhere in the range $1.5\times 10^6M_\odot\lap M_\bullet 
\lap 5\times 10^6M_\odot$ with equal likelihood 
\citep{valluri-04}.
Applying the $M_\bullet-\sigma$ relation \citep{fm-00}
gives $M_\bullet\approx 2.0\times 10^6M_\odot$
which we adopt here.
The black hole influence radius becomes $\sim 1.7$ pc
$\approx 0.5''$
and $T_r(r_h)\approx 2.2\times 10^{10}$ yr.
This is somewhat smaller than the estimate of $T_r(r_h)$ 
in the Galactic nucleus, suggesting that 
a collisional cusp could easily have been
regenerated in M32.
However, the cusp should only extend outward to 
$\sim 0.2 r_h\approx 0.1''$ at the distance
of M32, barely resolvable even with HST.
Depending on the exact value of $M_\bullet$,
the collisional cusp in M32 may or may not have been resolved.

Both the mass of the black hole in M31, and the dynamical
state of its nucleus, are less certain than  in the Milky
Way and M32 due to M31's complex morphology.
However most estimates of the relaxation time at the center of P2,
the nuclear component believed to contain the black hole, are
of order $10^{11}$ yr (e.g. Lauer et al. 1998).
M33 has a very short central relaxation time but
no dynamical signature of a black hole \citep{valluri-05}.

\subsection{Cusp Regeneration in Galaxies Beyond the Local Group}

Which galaxies beyond the Local Group should contain Bahcall-Wolf
cusps?
Figure~\ref{fig:tr} shows estimates of $T_r(r_h)$ in the
sample of early-type galaxies modelled by \cite{wang-04};
these are a subset of the \cite{magorrian-98} galaxies.
In this plot, filled symbols denote galaxies 
observed with angular resolution $\theta_{obs} < \theta_{r_h}$, 
and the size of the symbol is proportional to 
$\log(\theta_{r_h}/\theta_{obs})$.
(In galaxies where $r_h$ is not resolved, we note that 
the estimates of $T_r$ are particularly uncertain, since they
depend on an inward extrapolation of luminosity
profiles measured at $r>r_h$.)

Figure~\ref{fig:tr} reveals that the brightest spheroids, 
$M_V\lap -20$, have central relaxation times that always 
greatly exceed $10^{10}$ yr.
In these galaxies, a low-density core created by a 
binary supermassive black hole would persist
for the age of the universe.
However $T_r(r_h)$ drops with decreasing luminosity,
falling below $10^{10}$ yr for $M_V\gap -18$.
The Milky Way bulge falls on the relation defined
by the more distant galaxies, which is reassuring given the
uncertainties in its luminosity.
(We adopted a bulge blue absolute magnitude of $M_B=-17.6$
from \cite{marconi-03} and assumed $M_V=M_B-0.9$.)
However M32 appears to be shifted from the relation defined
by the other galaxies, as if it is the
dense core of a once much brighter galaxy.
This possibility has often been raised in the past
\citep{king-62,faber-73,nieto-87}.

\begin{figure}
\centering
\includegraphics[scale=0.45]{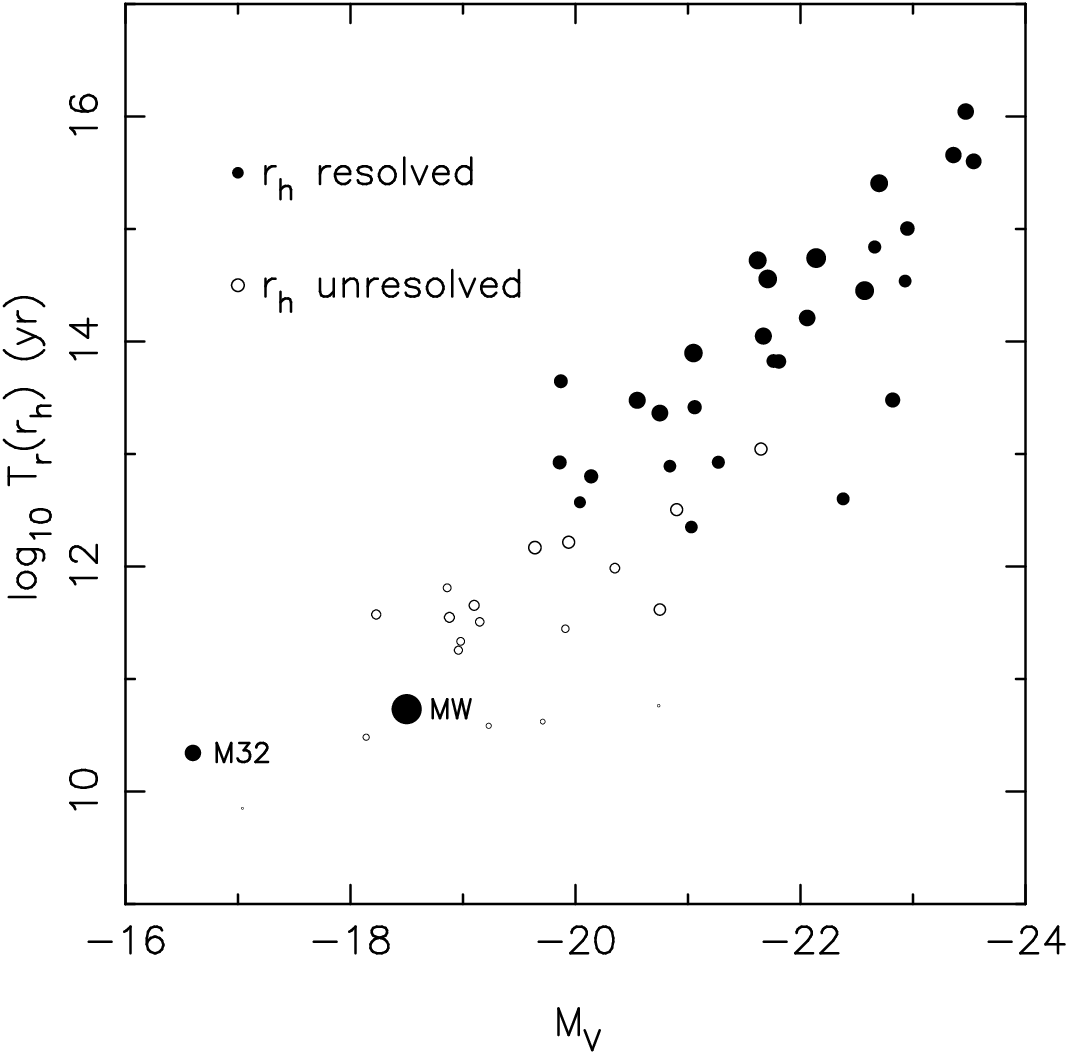}
\caption{Estimates of the relaxation time at the black hole's
influence radius, $r_h$,
in the sample of early-type galaxies modelled by \cite{wang-04}.
Black hole masses were computed from the $M_\bullet-\sigma$
relation \citep{mf-01b},
except in the case of the Milky Way, for which
$M_\bullet=3.7\times 10^6M_\odot$ was assumed
\citep{ghez-05}.
The stellar mass was set equal to $0.7M_\odot$ when computing
$T_r$.
Horizonal axis is absolute visual  magnitude of the
galaxy or, in the case of the Milky Way, the stellar bulge.
The size of the symbols is proportional to
$\log_{10}(\theta_{r_h}/\theta_{obs})$, where $\theta_{r_h}$ is
the angular size of the black hole's influence radius and
$\theta_{obs}$ is the observational resolution.
Filled symbols have $\theta_{r_h}>\theta_{obs}$ 
($r_h$ resolved) and open circles have
$\theta_{r_h}<\theta_{obs}$ ($r_h$ unresolved).
Values of $T_r(r_h)$ in the unresolved galaxies should
be considered approximate since the luminosity profiles
in these galaxies are not well known at $r<r_h$.
\label{fig:tr}
}
\end{figure}

Figure~\ref{fig:tr}, combined with the arguments in \S 5.1, 
suggests that spheroids fainter than $M_V\approx -18.5$ 
are dynamically old enough to contain Bahcall-Wolf cusps.

However outside of the Local Group, 
Figure~\ref{fig:tr} also suggests that
such cusps are unlikely to be resolved.
We can check this prediction by relating the angular size 
of the cusp to black hole mass via the $M_\bullet-\sigma$ relation,
$M_8\approx 1.66\sigma_{200}^{4.86}$ \citep{ff-05},
with $M_8\equiv M_\bullet/10^8M_\odot$ and 
$\sigma_{200}\equiv \sigma/200$ km s$^{-1}$,
and (temporarily) redefining $r_h$ as $GM_\bullet/\sigma^2\approx
11.2 M_8\sigma_{200}^{-2}\ \rm{pc}$.
Then
\beq
r_h\approx 2.76 M_8^{0.59}\ {\rm pc}.
\eeq
Taking for the outer radius of the cusp $r_0\approx 0.2 r_h$
(equation \ref{eq:spike}), its angular size becomes
\beq
\theta_0\approx 0.57'' M_8^{0.59}D_{Mpc}^{-1}
\label{eq:size}
\eeq
with $D_{Mpc}$ the distance to the galaxy in Mpc.
Thus, at the distance of the Virgo cluster, $\theta_0 > 0.1''$
implies $M_\bullet \gap 6\times 10^8M_\odot$;
however such massive black holes would almost certainly sit
in galaxies with central relaxation times 
longer than $10^{10}$ yr (Figure~\ref{fig:tr})
and a cusp would not have formed.
If we assume that black holes like the ones in
the Milky Way and M32 (i.e. $M_\bullet\approx 3\times 10^6M_\odot$)
are the most massive to be associated with collisionally-relaxed
nuclei, then the associated cusps
could be resolved to a distance of $\sim 0.7$ Mpc,
roughly the distance to M32 -- consistent with the
statement made above that the cusp in M32 is only
barely resolved.
Hence, collisional cusps are unlikely to be observed
in galaxies beyond the Local Group.

Unresolved density cusps might
appear as pointlike nuclei, particularly
in dE galaxies which have low central surface
brightnesses.
Pointlike nuclei are in fact nearly ubiquitous in elliptical
galaxies as faint as $M_V\approx -18$, disappearing for $M_V\gap -13$
\citep{vdb-86}.
Luminosities of the nuclei are observed to average $\sim 0.003$ 
times that of their host galaxies, albeit with considerable scatter
\citep{virgo-VIII}.
As shown in \S4, Bahcall-Wolf cusps entrain 
a mass of order $0.1M_\bullet$.
If the ratio of black hole mass to stellar mass that
characterizes bright galaxies, $M_\bullet/M_{gal}\approx 0.0013$
\citep{mf-01a},
also holds for dE galaxies, the luminosity associated
with  the cusps would be only $\sim 10^{-4}L_{gal}$,
too small to explain the majority of the observed nuclei.
On the other hand, essentially nothing is known about the
masses (or even the existence) of black holes in spheroids
fainter than $M_V=-18$ (with the exception of M32, probably
 a special case)
and it is possible that $M_\bullet \gap 10^{-3}M_{gal}$
in these galaxies.
It is intriguing to speculate that the disappearance of 
pointlike nuclei in dE galaxies fainter than $M_V\approx -13$ 
might signal the disappearance of the black holes.

We note that the nuclear cusps  of  the Milky Way and M32
extend approximately as power laws out to radii  far beyond
$r_h$.
Even these more extended cusps would be unresolved
beyond the Local Group and might
contain enough light to explain the pointlike nuclei.
The existence of these extended power-law cusps -- 
which presumably have little to do with the presence
of the black holes -- suggests that other mechanisms,
e.g. star formation triggered by gas infall, might
be as effective as stellar dynamics at re-generating
cusps.

The black holes in the Milky Way and M32 are among the smallest
with dynamically-determined masses \citep{ff-05}.
If smaller black holes do not exist, Figure~\ref{fig:tr}
suggests that Bahcall-Wolf cusps might be present only
in a small subset of spheroids containing black holes
with masses $10^6\msun\lap M_\bullet\lap 3\times 10^6\msun$.
However it has been argued that some late-type spirals
host AGN with black hole masses as low as $\sim 10^4M_\odot$
\citep{ho-04}.
If so, Figure~\ref{fig:tr} suggests that Bahcall-Wolf cusps
would be present around these black holes.

The presence of cores, or ``mass deficits,''  at the centers
of bright elliptical galaxies has been taken as evidence of
past merger events \citep{milos-02,ravin-02,graham-04}.
Mass deficits are observed to disappear in galaxies fainter than
$M_V\approx -19.5$ \citep{milos-02}.
Could this be due to cusp regeneration?
Figure~\ref{fig:tr} suggests an alternative explanation.
Galaxies fainter than $M_V= -19.5$ are mostly  unresolved on scales
of $r_h$, which is also the approximate size of a core
created by a binary supermassive black hole.
The lack of mass deficits in galaxies with
$M_V\gap -19.5$ probably just reflects a failure
to resolve the cores in these galaxies.

\subsection{Black Hole Feeding Rates}

The low-luminosity galaxies most likely to harbor
Bahcall-Wolf cusps (Figure~\ref{fig:tr}) are the same
galaxies that would dominate the overall 
rate of stellar tidal disruptions, assuming of course
that they contain black holes \citep{wang-04}.
Published estimates of $\dot N$, the rate of stellar disruptions,
in such low-luminosity galaxies \citep{su-99,mt-99,wang-04}
have almost always been based on an
inward extrapolation of luminosity profiles measured at $r>r_h$.
In principle, knowing that $\rho(r)$ has the Bahcall-Wolf
form near the black hole should allow a more accurate
estimate of $\dot N$ in the low-luminosity galaxies
that dominate the overall flaring rate.

Here we show that the presence of a Bahcall-Wolf
cusp implies a lower limit on $\dot N$, of order
$10^{-4}$ yr$^{-1}$.
The stellar density in the cusp is
\beq
\rho(r) \approx \rho(r_0)\left({r\over r_0}\right)^{-7/4}
\eeq
(equation \ref{eq:spike}), 
with $r_0=\alpha r_h$, $\alpha\approx 0.2$.
We can write $\rho(r_0)=KM_\bullet/r_h^3$, where
the constant $K$ depends on the form of $\rho(r)$
at $r>r_0$; assuming a $\rho\sim r^{-2}$ power
law for $r>r_0$, as in the Milky Way and many other
low-luminosity spheroids, we find $K\approx 4.0$.
The rate at which stars in the cusp
are fed to the black hole is approximately
\beq
\dot N_{cusp} \approx {4\pi\over m_\star} \int_0^{\alpha r_h}
{\rho\over T_r \ln(2/\theta_{lc})} r^2dr
\eeq
 \citep{ls-77,su-99}.
Here $\theta_{lc}\approx \sqrt{r_t/r}$ is the angular size
of the loss cone at radius $r$ and $r_t$ is the tidal
disruption radius, $r_t\approx (M_\bullet/m_\star)^{1/3}r_\star$.
This expression assumes that the feeding rate is limited
by diffusion, i.e. that the loss cone is ``empty'';
an equivalent statement is that $r_{crit}$, the 
radius above which a star can scatter in and out of the
loss cone in one orbital period, is greater than $\alpha r_h$.
In the case of the Milky Way black hole, it can
be shown that $0.2r_h<r_{crit}<r_h$.
Taking the slowly-varying logarithmic terms out of the
integral, we find a feeding rate for stars in the cusp:
\beq
\dot N_{cusp} \approx 1.6 {\ln\Lambda\over\ln(2/\theta_{lc})} 
\left({GM_\bullet\over r_h^3}\right)^{1/2}.
\eeq
Evaluating $\theta_{lc}$ at $r=\alpha r_h$
and setting $\ln\Lambda=15$, this becomes
\beq
\dot N_{cusp} \approx 7\times 10^{-5} {\rm yr}^{-1}
M_{\bullet, MW}^{1/2} r_{h,MW}^{-3/2}
\label{eq:ndot}
\eeq
where $M_{\bullet,MW}$ and $r_{h,MW}$ are in units of the
values quoted above for the Milky Way.
Thus, the flaring rate due to stars in the Milky Way cusp
is $\sim 10^{-4}$ yr$^{-1}$.
This is of course a lower limit on the total $\dot N$
since it ignores the contribution from stars outside the cusp,
at $r>\alpha r_0\approx 0.7$ pc.
In fainter spheroids, the $M_\bullet-\sigma$ relation,
combined with equation (\ref{eq:ndot}),
implies $\dot N \propto M_\bullet^{-0.4}$
and hence even higher flaring rates.

The Bahcall-Wolf solution will break down at 
radii where the physical collision time is shorter
than the diffusion time $\ln(2/\theta_{lc})T_r$.
Adopting the standard expression for the collision time,
\beq
T_{coll} = \left[16\sqrt{\pi}n\sigma r_\star^2(1+\Theta)\right]^{-1}
\eeq
with $\Theta\equiv Gm_\star/(2\sigma^2r_\star)$
and $n$ the number density of stars,
we find that physical collisions begin to affect the
stellar distribution at $r\lap 0.08$ pc $\lap 0.023 r_h$ for Solar-type
stars in the Galactic nucleus.

\subsection{Gravitational Lensing}

The central parts of galaxies can act as strong gravitational
lenses; the lack of a ``core'' image in observed lens systems
implies a lower limit on the stellar density
of the lensing galaxy within the central $\sim 10^2$ pc
\citep{rusin-01,keeton-03}.
Broken power-law density profiles like those in
equation (\ref{eq:spike}) have been used to
model lensing galaxies \citep{munoz-01,bowman-04}, although
the break radii in these studies were chosen to be much larger than
the value $r_0\approx 0.2 r_h$ that describes the Bahcall-Wolf
cusps (Figure~\ref{fig:all}).
However the presence or absence of the cusps
should have little effect on
the lensing properties of galaxies, because the mass 
contained within the cusp is small compared with $M_\bullet$, 
and because even the supermassive black holes contribute 
only slightly to the lensing signal \citep{rusin-05}.
The low-luminosity galaxies that are likely to contain
cusps (Figure~\ref{fig:tr}) are also unlikely to act
as lenses.

\subsection{Dark Matter}

The distribution of {\it dark matter} on sub-parsec scales
near the center of the Milky Way and other galaxies
is relevant to the so-called ``indirect detection''
problem, in  which inferences are drawn about the
properties of particle dark matter based on measurements
of its self-annihililation by-products \citep{mpla-05}.
A recent detection of TeV radiation from the Galactic
center by the HESS consortium \citep{aharonian-04} is 
consistent with a particle annihilation signal,
but only if the dark matter density in the inner few
parsecs is much higher than predicted
by an inward extrapolation of the standard, $\Lambda$CDM
halo models \citep{hooper-04}.
One possibility is that the dark matter forms a 
steep ``spike'' around the black  hole \citep{gs-99}.
Particle dark matter would not spontaneously form
a Bahcall-Wolf cusp since its
relaxation time is extremely long.
However, once a cusp forms in the {\it stars}, scattering
of dark matter particles off of stars would 
redistribute the  dark matter in phase space
on a time scale of order $T_r(r_h)$, the star-star
relaxation time \citep{merritt-04}.
The ultimate result is a $\rho \sim r^{-3/2}$ density cusp in the
dark matter \citep{gnedin-04}, 
but with possibly low normalization,
particularly if the  dark matter distribution was
previously modified by a binary black hole \citep{merritt-02}.
The $N$-body techniques applied here would be an effective
way to address this problem.

\acknowledgments
P. Cot\'e, L. Ferrarese, C. Keeton, S. Portegies Zwart,
A. Robinson, D. Rusin and the referee, H. Baumgardt, 
made comments and suggestions that improved this paper.
R. Genzel kindly provided the Galactic center number count 
data that are reproduced in Figure~\ref{fig:MW}.
This work was supported by grants 
AST-0071099, AST-0206031, AST-0420920 and AST-0437519 from the 
NSF, grant NNG04GJ48G from NASA,
and grant HST-AR-09519.01-A from STScI.


\begin{thebibliography}{}
\bibitem[Aarseth(1999)]{Aarseth:99} 
  Aarseth, S.~J.\ 1999, \pasp, 111, 1333 

\bibitem[Aarseth(2003)]{Aarseth:03} 
  Aarseth, S.~J.\ 2003, \apss, 285, 367 

\bibitem[Aharonian et al.(2004)]{aharonian-04} 
  Aharonian, F., et al.\ 2004, 
  \aap, 425, L1

\bibitem[Alexander(1999)]{alexander-99}
  Alexander, T. 1999,
  \apj, 527, 835

\bibitem[Bahcall \& Wolf(1976)]{bw-76}
  Bahcall, J. N. \& Wolf, R. A. 1976,
  \apj, 209, 214

\bibitem[Baumgardt et al.(2004)]{2004ApJ...613.1133B} 
  Baumgardt, H., 
  Makino, J., \& Ebisuzaki, T.\ 2004, \apj, 613, 1133 
 
\bibitem[Baumgardt et al.(2004)]{2004ApJ...613.1143B} 
  Baumgardt, H., 
  Makino, J., \& Ebisuzaki, T.\ 2004, \apj, 613, 1143 
 
\bibitem[Berczik, Merritt \& Spurzem(2005)]{bms-05}
  Berczik, P., Merritt, D. \& Spurzem, R. 2005,
  astro-ph/0507260 

\bibitem[Bertone \& Merritt(2005)]{mpla-05}
  Bertone, G., \& Merritt, D.\ 2005, 
  Modern Physics Letters A, 20, 1021 

\bibitem[Bowman et al.(2004)]{bowman-04} 
  Bowman, J.~D., Hewitt, J.~N., \& Kiger, J.~R.\ 2004, 
  \apj, 617, 81 
 
\bibitem[Boylan-Kolchin et al.(2004)]{boylan-04} 
  Boylan-Kolchin, M., Ma, C.-P., \& Quataert, E.\ 2004, 
  \apjl, 613, L37 
 
\bibitem[Cohn \& Kulsrud(1978)]{ck-78} 
	Cohn, H., \& Kulsrud, R. M. 1978, 
	\apj, 226, 1087

\bibitem[Cot\'e et al.(2005)]{virgo-VIII}
  Cot\'e, P., Piatek, S., Ferrarese, L., P., Jord\'an, A., Merritt, D.,
  Peng, E. W., Hasegan, M., Blakeslee, J. P., Mei, S., West, M. J.,
  Milosavljevi\'c, M., \& Tonry, J. L. 2005,
  submitted to The Astrophysical Journal.

\bibitem[Dehnen(1993)]{dehnen-93}
  Dehnen, W. 1993,
  \mnras, 265, 250 


\bibitem[Faber(1973)]{faber-73} 
	Faber, S. M. 1973, 
	\aj, 179, 423


\bibitem[Ferrarese et al.(2005)]{virgo-VI}
  Ferrarese, L. Cot\'e, P., Jord\'an, A., Peng, E. W., Blakeslee, J. P.,
  Piatek, S., Mei, S., Merritt, D., Milosavljevi\'c, M., Tonry, J. L.,
  \& West, M. J. 2005,
  submitted to The Astrophysical Journal.

\bibitem[Ferrarese \& Ford(2005)]{ff-05} 
  Ferrarese, L., \& Ford, H.\ 2005, 
  Space Science Reviews, 116, 523 

\bibitem[Ferrarese \& Merritt(2000)]{fm-00}
	Ferrarese, L., \& Merritt, D. 2000,
	\apj, 539, L9

\bibitem[Frank \& Rees(1976)]{fr-76} 
	Frank, J., \& Rees, M. J. 1976, 
	\mnras, 176, 633

\bibitem[Gebhardt et al.(1996)]{gebhardt-96}
  Gebhardt, K. et al. 1996,
  \aj, 112, 105

\bibitem[Genzel et al.(2003)]{genzel-03}
  Genzel, R., Schödel, R., Ott, T., Eisenhauer, F., Hofmann, R.,
  Lehnert, M., Eckart, A., Alexander, T., Sternberg, A., Lenzen, R., 
  Clénet, Y., Lacombe, F., Rouan, D., Renzini, A., \&
  Tacconi-Garman, L. E. 
  \apj, 594, 812

\bibitem[Ghez et al.(2005)]{ghez-05}
  Ghez, A. M., Salim, S., Hornstein, S. D., Tanner, A., Lu, J. R., 
  Morris, M., Becklin, E. E., \& Duch\^ene, G. 2005,
  \apj, 620, 744

\bibitem[Gnedin \& Primack(2004)]{gnedin-04} 
  Gnedin, O.~Y., \& Primack, J.~R.\ 2004, 
  Physical Review Letters, 93, 061302 
 
\bibitem[Gondolo \& Silk(1999)]{gs-99} 
  Gondolo, P., \& Silk, J.\ 1999, 
  Physical Review Letters, 83, 1719 
 
\bibitem[Graham(2004)]{graham-04}
  Graham, A. W. 2004,
  \apj, 613, L33

\bibitem[Hills(1983)]{hills-83}
	Hills, J. G. 1983,
	AJ, 88, 1269

\bibitem[Ho(2004)]{ho-04}
  Ho, L. 2004, in
	  Carnegie Observatories Astrophysics Series, Vol. 1: 
	  Coevolution of Black Holes and Galaxies, ed. L. C. Ho 
	  (Cambridge: Cambridge Univ. Press), p. 292

\bibitem[Hooper et al.(2004)]{hooper-04} 
  Hooper, D., de la Calle Perez, I., Silk, J., Ferrer, F., \& 
  Sarkar, S.\ 2004, 
  Journal of Cosmology and Astro-Particle Physics, 9, 2 
 
\bibitem[Keeton(2003)]{keeton-03} 
  Keeton, C.~R.\ 2003, 
  \apj, 582, 17 
 
\bibitem[King(1962)]{king-62}
  King, I. R. 1962,
  \aj, 67, 471

\bibitem[Lauer et al.(1998)]{lauer-98}
	Lauer, T. et al. 1998,
	AJ, 116, 2263

\bibitem[Lightman \& Shapiro(1977)]{ls-77}
  Lightman, A. P. \& Shapiro, S. L. 1977,
  \apj,  211, 244

\bibitem[Magorrian et al.(1998)]{magorrian-98} 
	Magorrian, J. et al. 1998, 
	\aj, 115, 2285

\bibitem[Magorrian \& Tremaine(1999)]{mt-99}
  Magorrian, J. \& Tremaine, S. 1999,
  \mnras, 309, 447

\bibitem[Marconi \& Hunt(2003)]{marconi-03}
  Marconi, A. \& Hunt, L. K. 2003, 
  \apj, 589, L21

\bibitem[Merritt(2004)]{merritt-04a}
	Merritt, D. 2004a,
	``Single and Binary Black Holes and their
	Influence on Nuclear Structure,'' in
	  Carnegie Observatories Astrophysics Series, Vol. 1: 
	  Coevolution of Black Holes and Galaxies, ed. L. C. Ho 
	  (Cambridge: Cambridge Univ. Press)

\bibitem[Merritt(2004)]{merritt-04} 
  Merritt, D.\ 2004, 
  Physical Review Letters, 92, 201304

\bibitem[Merritt \& Ferrarese(2001)]{mf-01a}
  Merritt, D. \& Ferrarese, L. 2001,
  \mnras, 320, L30

\bibitem[Merritt \& Ferrarese(2001)]{mf-01b}
  Merritt, D. \& Ferrarese, L. 2001,
  in ASP Conf. Ser. 249, The Central Kiloparsec of Starbursts
  and AGN: The La Palma Connection, ed. J. H. Knapen et al.
  (San Francisco: ASP), 335

\bibitem[Merritt, Mikkola \& Szell(2006)]{msm-06}
  Merritt, D., Mikkola, S. \& Szell, A. 2006,
  in preparation

\bibitem[Merritt \& Milosavljevic(2005)]{living-05}
  Merritt, D. \& Milosavljevic, M. 2005,
  ``Massive Black Hole Binary Evolution,''
  Living Reviews in Relativity

\bibitem[Merritt et al.(2004b)]{merritt-04b}
  Merritt, D., Milosavljevic, M., Favata, M., Hughes, S. A. \& Holz, D.E. 2004,
  \apj, 607, L9 

\bibitem[Merritt et al.(2002)]{merritt-02}
  Merritt, D., Milosavljevic, M., Verde, L. \& Jimenez, R. 2002,
  \prl, 88, 191301

\bibitem[Merritt \& Tremblay(1994)]{mt-94}
  Merritt, D. \& Tremblay, B. 1994,
  \aj, 108, 514

\bibitem[Merritt \& Wang(2005)]{wang-05}
  Merritt, D. \& Wang, J. 2005,
  \apj, 621, L101

\bibitem[Mikkola \& Aarseth(1990)]{MA:90} 
  Mikkola, S., \& Aarseth, S.~J.\ 1990, 
  Celestial Mechanics and Dynamical Astronomy, 47, 375

\bibitem[Mikkola \& Aarseth(1993)]{MA:93} 
  Mikkola, S., \& Aarseth, S.~J.\ 1993, 
  Celestial Mechanics and Dynamical Astronomy, 57, 439

\bibitem[Mikkola \& Valtonen(1992)]{mv-92} 
  Mikkola, S., \& Valtonen, M. J.\ 1992, 
  \mnras, 259, 115

\bibitem[Milosavljevic \& Meritt(2001)]{mm-01}
	Milosavljevic, M. \& Merritt, D. 2001,
	\apj, 563, 34

\bibitem[Milosavljevic \& Meritt(2003)]{mm-03}
	Milosavljevic, M. \& Merritt, D. 2003,
	\apj, 596, 860

\bibitem[Milosavljevic et al.(2002)]{milos-02}
  Milosavljevic, M., Merritt, D., Rest, A., \& van den Bosch, F. C. 2002,
  \mnras, 331, 51

\bibitem[Mu{\~n}oz et al.(2001)]{munoz-01} 
  Mu{\~n}oz, J.~A., Kochanek, C.~S., \& Keeton, C.~R.\ 2001, 
  \apj, 558, 657 

\bibitem[Murphy, Cohn \& Durisen(1991)]{murphy-91}
  Murphy, B. W., Cohn, H. N. \& Durisen, R. H. 1991,
  \apj, 370, 60

\bibitem[Nakano \& Makino(1999a)]{nakano-99a}
  Nakano, T. \& Makino, J. 1999a,
  \apj, 510, 155

\bibitem[Nakano \& Makino(1999b)]{nakano-99b}
  Nakano, T. \& Makino, J. 1999b,
  \apj, 525, L77

\bibitem[Nieto \& Prugniel(1987)]{nieto-87}
  Nieto, J.-L. \& Prugniel, P. 1987,
  \aa, 186, 30

\bibitem[Preto, Merritt \& Spurzem(2004)]{preto-04}
  Preto, M., Merritt, D. \& Spurzem, R. 2004,
  \apj, 613, L109

\bibitem[Quinlan(1996)]{quinlan-96}
  Quinlan, G. D. 1996,
  New Astron. 1, 35

\bibitem[Quinlan \& Hernquist(1997)]{quinlan-97}
  Quinlan, G. D. 1997,
  New Astron. 2, 533

\bibitem[Ravindranath, Ho \& Filippenko(2002)]{ravin-02}
  Ravindranath, S., Ho, L. C. \& Filippenko, A. V. 2002,
  \apj, 566, 801

\bibitem[Rest et al.(2001)]{rest-01}
  Rest, A. et al. 2001,
  \aj, 121, 2431

\bibitem[Rusin et al.(2005)]{rusin-05} 
  Rusin, D., Keeton, C.~R., \& Winn, J.~N.\ 2005, 
  \apjl, 627, L9


\bibitem[Rusin \& Ma(2001)]{rusin-01} 
  Rusin, D., \& Ma, C.-P.\ 2001, 
  \apjl, 549, L33 

\bibitem[Saslaw, Valtonen \& Aarseth(1974)]{saslaw-74}
  Saslaw, W. C., Valtonen, M. J., \& Aarseth, S. J. 1974,
  \apj, 190, 253

\bibitem[Spitzer(1987)]{spitzer-87} 
  Spitzer, L.\ 1987, 
  ``Dynamical Evolution of Globular Clusters''
  (Princeton: Princeton University Press)

\bibitem[Syer \& Ulmer(1999)]{su-99}
  Syer, D. \& Ulmer, A. 1999,
  \mnras, 306, 35

\bibitem[Szell, Merritt \& Mikkola(2005)]{smm-05}
  Szell, A., Merritt, D. \& Mikkola, S. 2005,
  in Nonlinear Dynamics in Astronomy \& Physics,
  ed. S. T. Gottesman, J.-R. Buchler \& M. E. Mahon,
  Ann. N. Y. Acad. Sci. 1045, 225

\bibitem[Valluri et al.(2005)]{valluri-05}
  Valluri, M., Ferrarese, L., Merritt, D. \& Joseph, C. L. 2005,
  \apj, 628, 137

\bibitem[Valluri, Merritt \& Emsellem(2004)]{valluri-04}
  Valluri, M., Merritt, D., \& Emsellem, E. 2004,
  \apj, 602, 66

\bibitem[van den Bergh(1986)]{vdb-86}
  van den Bergh, S. 1986,
  \aj, 91, 271

\bibitem[Volonteri et al.(2003)]{volonteri-03} 
  Volonteri, M., Madau, P., \& Haardt, F.\ 2003, 
  \apj, 593, 661 

\bibitem[Wang \& Merritt(2004)]{wang-04}
  Wang, J. \& Merritt, D. 2004,
  \apj, 600, 149 

\bibitem[Young(1977)]{young-77}
	Young, P. J. 1977,
	\apj, 215, 36

\bibitem[Yu(2002)]{2002MNRAS.331..935Y} 
  Yu, Q.\ 2002, \mnras, 331, 935 
 

\end{thebibliography}
\end{document}